\documentclass[10pt, conference]{IEEEtran}
\IEEEoverridecommandlockouts
\usepackage{cite}
\usepackage{amsmath,amssymb,amsfonts,amsthm}
\usepackage{algorithmic}
\usepackage{graphicx}
\usepackage{textcomp}
\usepackage{xcolor}
\usepackage{comment}
\theoremstyle{definition}
\newtheorem{remark}{Remark}
\newtheorem{assumption}{Assumption}

\begin{document}

\title{Dual-Security for Indoor OFDM-ISAC Systems via Temporal Artificial Noise}
\author{
\IEEEauthorblockN{Yinchao Yang, Yathreb Bouazizi, Prabhat Raj Gautam, Michael Breza, and Julie A. McCann}
\IEEEauthorblockA{Imperial College London\\
Emails: \{yinchao.yang, y.bouazizi18, p.gautam, michael.breza04, j.mccann\}@imperial.ac.uk}
\thanks{This work was supported by the EPSRC through the Communications Hub for Empowering Distributed Cloud Computing Applications and Research (CHEDDAR) [grant numbers EP/X040518/1 and EP/Y037421/1], IntentMAPs [EP/Z00084X/1], and the National Edge AI Hub [EP/Y028813/1].}
}

\maketitle

\begin{abstract}
    With the rapid development of integrated sensing and communication (ISAC) as a key enabler for future wireless networks, ensuring the security of both communication and sensing functions has become increasingly important. Current secure ISAC studies focus restrictively either on the communication or the sensing security, but not both. To bridge this gap, this paper investigates security for both, i.e., dual-security, in indoor orthogonal frequency division multiplexing (OFDM) based ISAC systems. Specifically, we consider a scenario in which a sensing user (SU) is authorized for sensing but may eavesdrop on communication data, while a communication user (CU) is authorized for communication but may perform unauthorized sensing. We chose this scenario as the pathological case where an authorized eavesdropper has more information and is more effective than an unauthorized one. To address this case, we propose the use of temporal artificial noise (AN) to prevent malicious CU sensing by enlarging its time-domain sensing error, and simultaneously degrade SU data eavesdropping by reducing its frequency-domain signal-to-interference-plus-noise ratio (SINR) with standard OFDM receiver processing. Meanwhile, our proposed scheme guarantees the sensing performance of the SU and the communication performance of the CU. We present numerical results that demonstrate AN can effectively provide dual protection for sensing and communication in OFDM-ISAC systems while guaranteeing the performance of legitimate users.
\end{abstract}


\section{Introduction}

Integrated sensing and communication (ISAC) has emerged as a key enabling technology for future wireless systems, owing to its ability to support sensing and communication over a shared hardware platform, spectrum, waveform, and other system resources \cite{zhang2026integrated}. ISAC supports a wide range of applications, spanning outdoor scenarios such as traffic monitoring and management to indoor scenarios such as person positioning and human motion recognition \cite{etsi1integrated}. Although indoor and outdoor ISAC applications both can use communication-centric orthogonal frequency division multiplexing (OFDM) waveforms, they have very different system parameters and requirements \cite{liu2018mu}. Outdoor ISAC systems share key characteristics with 5G cellular networks, including the use of large-scale multiple-input multiple-output (MIMO) antenna arrays, a large number of subcarriers and symbols, and often use range-Doppler maps for sensing. Indoor, WiFi-based ISAC systems have a small number of antennas (single or $2\!\times\!2$ MIMO), subcarriers and symbols, and use channel characteristics like channel state information (CSI) or received signal strength indicator (RSSI) for sensing \cite{hernandez2022wifi}. 

Despite the promises of ISAC, the integration of communication and sensing creates new security challenges \cite{su2025integrating}. Communication security, also known as data security, entails protecting the data from being decoded by malicious actors. Sensing security involves protecting targets from being detected or tracked by unintended receivers using the ISAC signals reflected from the targets.
Dual-security refers to the joint protection of both communication and sensing functionalities. Current secure ISAC studies focus either on the communication or the sensing security, but not both. For data security, survey papers \cite{han2025next, niu2025survey} review a range of physical layer security (PLS) techniques, which provide a complementary and information-theoretic layer of protection alongside conventional encryption methods. More recently, several works \cite{zou2024securing, han2025sensing, du2025securing, chen2025sensing, yang2026dual} have begun to investigate sensing security in outdoor ISAC scenarios. However, dual-security and indoor scenarios remain largely unexplored.

The security concern for indoor sensing differs from that of outdoor sensing, which primarily focuses on range-Doppler information. The CSI used for indoor ISAC contains personal identifiable information, such as vital signs, that can be used for covert surveillance, which violates data protection laws, e.g., EU GDPR Article 9 \cite{gdpr_article9}, and US BIPA Act \cite{BIPA_2008}. Existing secure indoor sensing designs, such as \cite{ghiro2022implementation, naghibi2011mimo}, do not explicitly guarantee communication performance and are therefore not applicable to ISAC systems.

In this paper, we address dual-security for indoor ISAC systems and propose a temporal (i.e., time-domain) artificial noise (AN) design for OFDM-ISAC systems to secure both sensing and communication. We consider two types of users, as specified in the ETSI report \cite{ETSI_GR_ISC_004}. A sensing user (SU) is a cooperative/active bistatic receiver using data payload for sensing \cite{li2026rethinking}, but is not authorized to use the communication service. An authorized communication user (CU) communicates but is not authorized to use the sensing service. Both CU and SU can maliciously capture data from services for which they are not authorized. We design a temporal AN that is embedded in the cyclic prefix (CP) and designed to be null at the CU, thereby preserving its communication performance. Since the AN is not null at the SU, the SU suffers a degraded frequency-domain signal-to-interference-plus-noise ratio (SINR), which ensures communication security. Meanwhile, the proposed design allows both the target echo and the SU's reference signal to contain the same temporal AN, enabling accurate time-domain estimation of the target response channel at the SU. In contrast, because the CU does not receive the temporal AN, the resulting mismatch degrades its sensing performance, thereby ensuring sensing security. Our main contributions are:
\begin{itemize}
    \item We consider a new dual-threat model in an ISAC system, where the SU is authorized for sensing services but acts as a potential communication eavesdropper (Eve), while the CU is authorized for communication services but acts as a sensing Eve.
    
    \item We propose a novel temporal AN design that leverages the CP and exploits the duality between the time and frequency domains to guarantee legitimate service quality and suppress unauthorized eavesdropping simultaneously.
\end{itemize}

\section{System Model}

\begin{figure}[t]
    \centering
    \includegraphics[width=0.75\columnwidth]{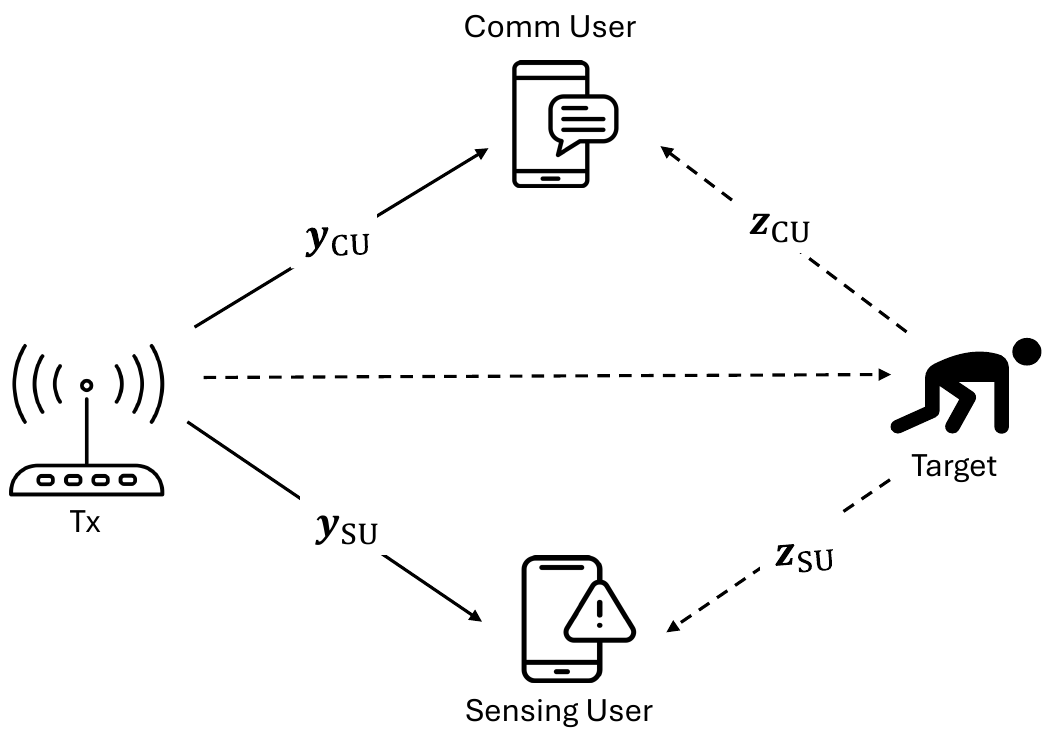}
    \caption{An illustration of the considered ISAC system model with one Tx, one CU, one SU, and one target. Solid lines represent communication links, while dashed lines represent sensing links.}
    \label{sys model}
\end{figure}

As illustrated in Fig.~\ref{sys model}, our system consists of an indoor single-antenna OFDM-based ISAC system with one transmitter (Tx), one CU, one SU, and one target. We consider eavesdropping attacks that come from authorized users who are part of the network and are synchronized. Both types of users can maliciously capture data from services for which they are not authorized. For the remainder of this paper, we consider only the case of threats from authorized users, but our solution generalizes to the easier case of threats from unauthorized users, i.e., Eves that have not subscribed to any service.
The assumptions used in this paper are listed below:
\begin{assumption}\label{assump 1}
    The Tx is assumed to have perfect knowledge of each user's service type (communication or sensing) and the downlink CSI. 
\end{assumption}

\begin{assumption}\label{assump 2}
    The wireless channel is assumed to be quasi-static and remains constant within each coherence block.
\end{assumption}

\begin{assumption}\label{assump 3}
    The communication and echo signals are assumed to be temporally separated to avoid interference, occupying distinct time slots within each coherence block. 
\end{assumption}

Assumption~\ref{assump 1} can be satisfied through an initial uplink-downlink channel estimation procedure, as all users are subscribed to services. Assumption~\ref{assump 2} is a commonly used assumption, which enables the analysis to focus on a single OFDM symbol rather than multiple ones, thereby simplifying the derivation. Assumption~\ref{assump 3} enables the analysis to focus on a single antenna system for analytical clarity, where the echo signal arrives later than the communication signal due to the two-way propagation path. 

\begin{figure}[!t]
    \centering
    \includegraphics[width=0.75\columnwidth]{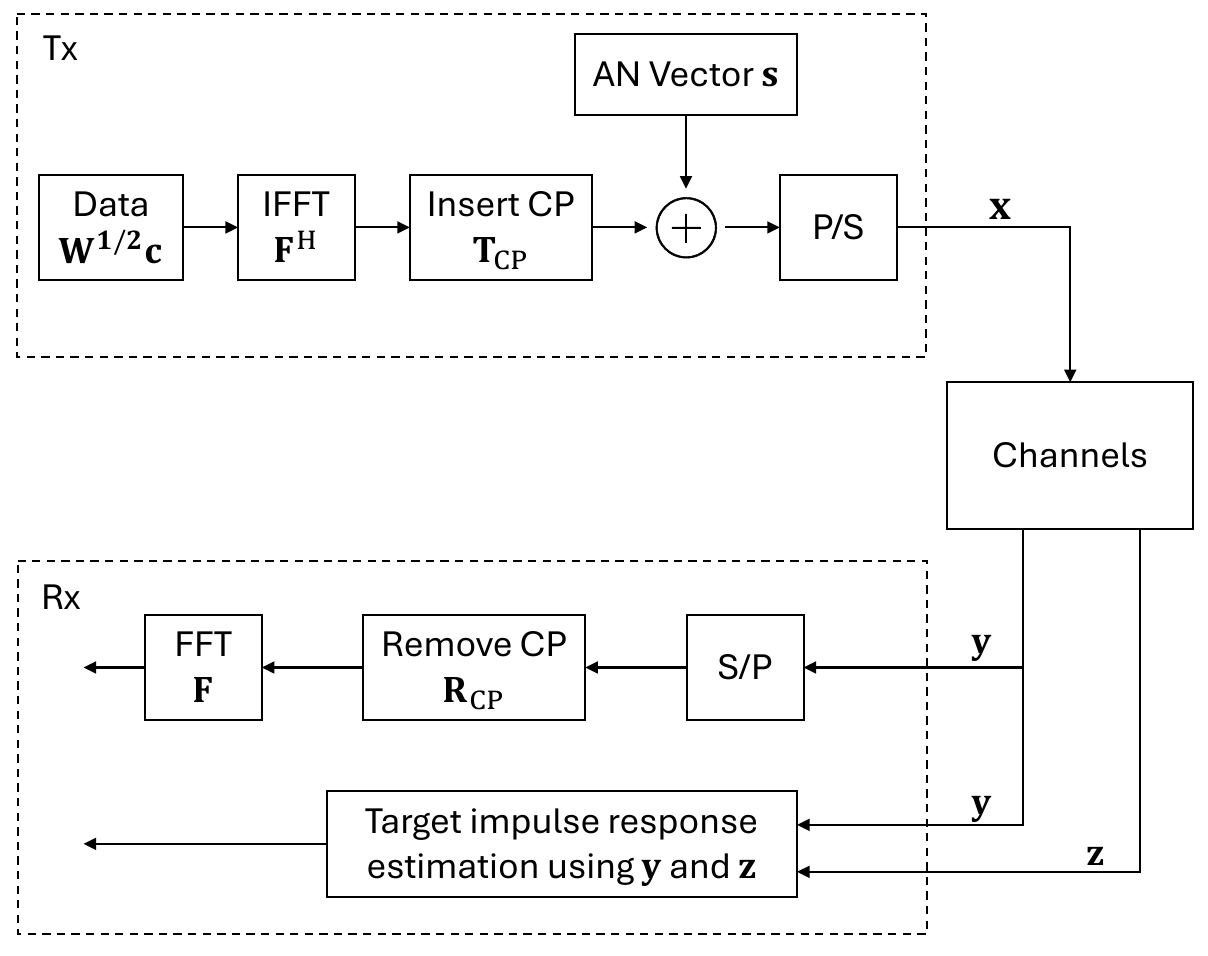}
    \caption{A block diagram of the considered Tx and receiver (Rx) signal models. The receiver (Rx) may represent either the CU or the SU, as both share the same structure. The channel block includes both communication and sensing channels with AWGN. The parallel-to-serial (P/S) and serial-to-parallel (S/P) conversions are omitted from the analysis, since the focus is on per OFDM symbol processing. From the CU perspective, the aims are to ensure high-quality FFT output while degrading the target response estimation output. From the SU perspective, the objective is the opposite.}
    \label{DSP fig}
\end{figure}

Finally, Fig.~\ref{DSP fig} provides an overview of the considered signal models and processing steps, and each component is described in detail in the following subsections.

\subsection{OFDM Signal}

Let $N_{c}$ and $N_{cp}$ denote the number of subcarriers and the CP length, respectively. The discrete time-domain OFDM signal $\mathbf{x} \in \mathbb{C}^{(N_{c} + N_{cp}) \times 1}$ is expressed as:
\begin{equation}
    \mathbf{x} = \mathbf{T}_{\mathrm{CP}} \mathbf{F}^{\mathrm{H}} \mathbf{W}^{1/2} \mathbf{c} + \mathbf{s} \triangleq \tilde{\mathbf{w}} + \mathbf{s},
\end{equation}
where $\mathbf{T}_{\mathrm{CP}} \in \mathbb{C}^{(N_{c} + N_{cp}) \times N_{c}}$ is the CP insertion matrix \cite{qin2013power}, $\mathbf{F}^{\mathrm{H}}$ is the $N_{c}$-point inverse fast Fourier transform (IFFT) matrix, and $\mathbf{W}^{1/2} = \mathrm{diag}(\sqrt{w_{1}}, \dots, \sqrt{w_{N_c}})$ is the power allocation matrix. The data vector $\mathbf{c}$ contains i.i.d. symbols with $\mathbb{E}[\mathbf{c}\mathbf{c}^{\mathrm{H}}] = \mathbf{I}_{N_{c}}$. To enhance PLS, an AN vector $\mathbf{s} = \mathbf{U} \mathbf{v}$ is superimposed, where $\mathbf{U}$ is a semi-unitary matrix ($\mathbf{U}^{\mathrm{H}} \mathbf{U} = \mathbf{I}_{N_{cp}}$), and $\mathbf{v} \sim \mathcal{CN}(\mathbf{0}, \boldsymbol{\Sigma})$ with $\boldsymbol{\Sigma}$ being a diagonal matrix \cite{marzban2018securing}. The data symbols $\mathbf{c}$ and the AN vector $\mathbf{s}$ are assumed to be statistically independent \cite{niu2025survey}, as such, we denote the covariance matrix $\mathbf{R}_{x} = \mathbf{R}_{\tilde{w}} + \mathbf{U} \boldsymbol{\Sigma} \mathbf{U}^{\mathrm{H}}$.

\subsection{Communication User Models}

At the CU side, the CP is first discarded, followed by an $N_c$ point fast Fourier transform (FFT) to transition the signal back to the frequency domain. The processed signal $\mathbf{y}_{\mathrm{CU}} \in \mathbb{C}^{N_{c} \times 1}$ is expressed as:
\begin{align}\label{Comm_Rx}
    \mathbf{y}_{\mathrm{CU}} &= \mathbf{F} \mathbf{R}_{\mathrm{CP}} \mathbf{H}_{\mathrm{CU}} \mathbf{x} + \mathbf{n} \notag \\
    & = \tilde{\mathbf{H}}_{\mathrm{CU}}\mathbf{W}^{1/2} \mathbf{c} + \mathbf{F} \mathbf{R}_{\mathrm{CP}} \mathbf{H}_{\mathrm{CU}} \mathbf{U} \mathbf{v} + \mathbf{n} \notag \\
    & \overset{(a)}{=} \tilde{\mathbf{H}}_{\mathrm{CU}} \mathbf{W}^{1/2} \mathbf{c} + \mathbf{n},
\end{align}
where $\mathbf{R}_{\mathrm{CP}} \in \mathbb{C}^{N_c \times (N_{c} + N_{cp})}$ denotes the CP removal matrix and $\mathbf{F}\in \mathbb{C}^{N_{c} \times N_c}$ is the unitary FFT matrix. 
The matrix $\mathbf{H}_{\mathrm{CU}} \in \mathbb{C}^{(N_{c} + N_{cp}) \times (N_{c} + N_{cp})}$ represents the Toeplitz (Toep) structured channel impulse response (CIR) matrix, and $\tilde{\mathbf{H}}_{\mathrm{CU}}~\in~\mathbb{C}^{N_{c} \times N_{c}}$ is the diagonal frequency-domain channel matrix. 
The term $\mathbf{n}$ represents the processed additive white Gaussian noise (AWGN) with $\mathbf{n} \sim \mathcal{CN}(\mathbf{0}, \sigma^{2} \mathbf{I}_{N_c})$. 
In \eqref{Comm_Rx}, equality $(a)$ holds under the design constraint that the AN lies in the null space of the communication channel, i.e., $\mathbf{R}_{\mathrm{CP}} \mathbf{H}_{\mathrm{CU}} \mathbf{U} = \mathbf{0}$. 
Consequently, the columns of $\mathbf{U}$ are chosen from the basis of $\mathrm{null}(\mathbf{R}_{\mathrm{CP}} \mathbf{H}_{\mathrm{CU}})$.

Simultaneously, the CU acts as a potential sensing Eve attempting to perform target sensing using the received echo. The received time-domain echo signal $\mathbf{z}_{\mathrm{CU}}~\in~\mathbb{C}^{(N_{cp} + N_c) \times 1}$ at the CU is modeled as:
\begin{equation}\label{Rx_echo}
\begin{aligned}
    \mathbf{z}_{\mathrm{CU}} = \mathbf{X} \mathbf{j}_{\mathrm{CU}} + \mathbf{n}_{s},
\end{aligned}
\end{equation}
where $\mathbf{X} \in \mathbb{C}^{(N_{c} + N_{cp}) \times (N_{c} + N_{cp})}$ is the Toep matrix constructed from $\mathbf{x}$ \cite{chen2025rethinking}, and $\mathbf{j}_{\mathrm{CU}} \in \mathbb{C}^{(N_{c} + N_{cp}) \times 1}$ is the target impulse response vector whose first $L$ elements are non-zero\footnote{Although $\mathbf{X}$ can be structured with size $(N_c + N_{cp}) \times L$, this zero-padding formulation ensures $\mathbf{X}$ is a full-rank square matrix, thereby facilitating direct inversion and avoiding the Moore-Penrose pseudo inverse.}. The channel $\mathbf{j}_{\mathrm{CU}}$ is assumed to be complex Gaussian distributed with zero mean and covariance of $\sigma^2_{j} [\mathbf{I}_{L} \; \mathbf{0}_{N_c + N_{cp} - L}]$ \cite{liu2014channel}. The term $\mathbf{n}_{s}~\sim~\mathcal{CN}(\mathbf{0}, \sigma_s^2 \mathbf{I}_{N_c + N_{cp}})$ denotes the AWGN.

\subsection{Sensing User Models}

The SU is modeled as a cooperative bistatic receiver and utilizes the entire transmitted frame as a reference signal for sensing. The direct-path time-domain reference signal $\mathbf{y}_{\mathrm{SU, ref}} \in \mathbb{C}^{(N_{cp} + N_c) \times 1}$ is modeled as:
\begin{equation}\label{SU ref}
    \mathbf{y}_{\mathrm{SU, ref}} = \mathbf{H}_{\mathrm{SU}} \mathbf{x} + \mathbf{n},
\end{equation}
where $\mathbf{H}_{\mathrm{SU}}$ denotes the direct-link Toep channel between the transmitter and the SU.

The received time-domain echo signal $\mathbf{z}_{\mathrm{SU}} \in \mathbb{C}^{(N_{cp} + N_c) \times 1}$ is expressed as:
\begin{equation}\label{SU echo}
    \mathbf{z}_{\mathrm{SU}} = \mathbf{X} \mathbf{j}_{\mathrm{SU}} + \mathbf{n}_s,
\end{equation}
where $\mathbf{j}_{\mathrm{SU}}~\sim~\mathcal{CN}(\mathbf{0}, \sigma_{j}^2 \mathbf{I}_{N_c + N_{cp}})$ represents the target impulse response relative to the SU.

In the event that the SU attempts to illicitly decode the communication message from the reference signal \eqref{SU ref}, it must process the signal similarly to the CU. By removing the CP and applying an FFT, the frequency-domain signal at the SU $\mathbf{y}_{\mathrm{SU, decode}} \in \mathbb{C}^{N_c \times 1}$ becomes:
\begin{align} 
    \mathbf{y}_{\mathrm{SU, decode}} &= \mathbf{F} \mathbf{R}_{\mathrm{CP}} \mathbf{H}_{\mathrm{SU}} \mathbf{x} + \mathbf{n} \notag \\
    & = \tilde{\mathbf{H}}_{\mathrm{SU}}\mathbf{W}^{1/2} \mathbf{c} + \mathbf{F} \mathbf{R}_{\mathrm{CP}} \mathbf{H}_{\mathrm{SU}} \mathbf{U} \mathbf{v} + \mathbf{n},
\end{align}
where the term $\mathbf{F} \mathbf{R}_{\mathrm{CP}} \mathbf{H}_{\mathrm{SU}} \mathbf{U} \mathbf{v}$ represents the AN. Since $\mathbf{U}$ is designed based on the CU channel $\mathbf{H}_{\mathrm{CU}}$, it generally does not fall into the null space of $\mathbf{H}_{\mathrm{SU}}$, thereby providing physical layer protection for the confidential data. Additionally, $\tilde{\mathbf{H}}_{\mathrm{SU}}~\in~\mathbb{C}^{N_{c} \times N_{c}}$ is the diagonal frequency-domain channel matrix, and $\mathbf{n}$ is AWGN.

\section{Performance Metrics}

This section formulates the performance metrics used to evaluate the proposed design. Specifically, we characterize the legitimate communication and sensing performance of the CU and SU, respectively, as well as the corresponding security-related metrics for unauthorized communication decoding and sensing.

\subsection{Communication User}

\subsubsection{Subscribed Service}
To ensure the quality-of-service (QoS) for the CU, the signal-to-noise ratio (SNR) is formulated based on \eqref{Comm_Rx}:
\begin{equation}\label{cu snr}
    \gamma_{\mathrm{CU}} = \frac{1}{\sigma^2} \mathrm{Tr} \left( \Tilde{\mathbf{H}}_{\mathrm{CU}} \mathbf{W} \Tilde{\mathbf{H}}_{\mathrm{CU}}^{\mathrm{H}} \right),
\end{equation}
where the AN interference is eliminated at the CU through the null-space projection $\mathbf{U} \in \mathrm{null}(\mathbf{R}_{\mathrm{CP}} \mathbf{H}_{\mathrm{CU}})$.

\subsubsection{Unauthorized Service}

As a potential sensing Eve, the CU does not receive the AN signal $\mathbf{s}$. Consequently, it can only use the information-bearing signal as a reference signal. Let $\tilde{\mathbf{W}} = \mathrm{Toep}(\tilde{\mathbf{w}})$ denote the perceived time-domain reference signal matrix. The CU least squares estimate (LSE) of the target channel is:
\begin{equation}
    \hat{\mathbf{j}}_{\mathrm{CU}} = \tilde{\mathbf{W}}^{-1} \mathbf{z}_{\mathrm{CU}} = \mathbf{j}_{\mathrm{CU}} + \tilde{\mathbf{W}}^{-1} \left( \mathrm{Toep}(\mathbf{s}) \mathbf{j}_{\mathrm{CU}} + \mathbf{n}_{s} \right).
\end{equation}

The estimation error comprises the AN-induced interference and AWGN\footnote{A strong adversary may iteratively estimate both the channel and the AN. However, channel estimation errors may propagate to AN estimation and degrade its accuracy. Such advanced adversarial capabilities are beyond the scope of this work.}. The resulting mean squared error (MSE) is derived as:
\begin{align}\label{cu Sensing error}
        \epsilon_{\mathrm{CU}} &= \mathbb{E}\left[  \left| \tilde{\mathbf{W}}^{-1} \left( \mathrm{Toep}\left( \mathbf{s}\right) \mathbf{j}_{\mathrm{CU}}  + \mathbf{n}_{s} \right)  \right|^{2} \right] \notag \\
        & = \mathbb{E}\left[ \mathrm{Tr} \left( \tilde{\mathbf{W}}^{-1}  \mathrm{Toep}\left( \mathbf{s}\right) \mathbf{j}_{\mathrm{CU}} \mathbf{j}_{\mathrm{CU}}^{\mathrm{H}} \mathrm{Toep}\left( \mathbf{s}\right)^{\mathrm{H}} (\tilde{\mathbf{W}}^{-1})^{\mathrm{H}}  \right) \right] \notag \\
        & \quad +  \mathbb{E}\left[ \mathrm{Tr} \left( \tilde{\mathbf{W}}^{-1}  \mathbf{n}_{s}\mathbf{n}_{s}^{\mathrm{H}}(\tilde{\mathbf{W}}^{-1})^{\mathrm{H}}  \right) \right] \notag \\
        & \overset{(a)}{\geq} \mathrm{Tr} \left( \left( \sigma^2_{j} \mathbf{Q} \mathbf{U} \boldsymbol{\Sigma} \mathbf{U}^{\mathrm{H}}  \mathbf{Q}^{\mathrm{H}} + \sigma^2_{s} \mathbf{I} \right) \left( \mathbf{Q} \mathbf{R}_{\tilde{w}} \mathbf{Q}^{\mathrm{H}} \right) ^{-1} \right),
\end{align}
where $\sigma_j^2$ is the variance of the target response channel that can be observed over time, and the inequality $(a)$ comes from Jensen's inequality. The diagonal matrix $\mathbf{Q}$ accounts for the power distribution of the Toep structure, where elements $q_{i} = \sqrt{L}$ for $1 \le i \le N_c + N_{cp} - L + 1$, and decreases linearly to $1$ for the remaining indices.

To illustrate the construction of $\mathbf{Q}$, consider an $L=3$ tap channel and $\mathbf{s} = [s_1,\ldots,s_5]^{\mathrm T}$, thus, the Toep matrix is:
\begin{equation}
\operatorname{Toep}(\mathbf{s})
=
\begin{bmatrix}
s_1 & 0   & 0 \\
s_2 & s_1 & 0 \\
s_3 & s_2 & s_1 \\
s_4 & s_3 & s_2 \\
s_5 & s_4 & s_3
\end{bmatrix}.
\nonumber
\end{equation}

With $\mathbb{E}[\mathbf{j}_{\mathrm{CU}}\mathbf{j}_{\mathrm{CU}}^{\mathrm H}]
=\sigma_j^2\mathbf{I}$, we have the following
\begin{align}
&\mathbb{E}\!\left[
\operatorname{Tr}\left(
\operatorname{Toep}(\mathbf{s})
\mathbf{j}_{\mathrm{CU}}\mathbf{j}_{\mathrm{CU}}^{\mathrm H}
\operatorname{Toep}(\mathbf{s})^{\mathrm H}
\right)
\right]
\nonumber\\
&=\sigma_j^2
\mathbb{E}\!\left[
\operatorname{Tr}\left(
\operatorname{Toep}(\mathbf{s})
\operatorname{Toep}(\mathbf{s})^{\mathrm H}
\right)
\right]
\nonumber\\
&=\sigma_j^2
\mathbb{E}\!\left[
3|s_1|^2+3|s_2|^2+3|s_3|^2
+2|s_4|^2+|s_5|^2
\right]
\nonumber\\
&=\sigma_j^2
\operatorname{Tr}\left(
\mathbf{Q}
\mathbb{E}\!\left[\mathbf{s}\mathbf{s}^{\mathrm H}\right]
\mathbf{Q}^{\mathrm H}
\right) = \sigma_j^2
\operatorname{Tr}\left(
\mathbf{Q}
\mathbf{U}\boldsymbol{\Sigma}\mathbf{U}^{\mathrm H}
\mathbf{Q}^{\mathrm H}
\right),
\nonumber
\end{align}
where $\mathbf{Q} = \operatorname{diag}\left( \sqrt{3},\sqrt{3},\sqrt{3},\sqrt{2},1 \right)$.

\subsection{Sensing User}

\subsubsection{Subscribed Service}
While the SU could theoretically employ Bayesian estimators given its cooperative status, we utilize the LSE to establish a performance lower bound, representing a worst-case estimation performance. By assuming the SU can perfectly recover $\mathbf{X}$ from $\mathbf{y}_{\mathrm{SU, ref}}$, the target channel can be estimated via:
\begin{equation}
\hat{\mathbf{j}}_{\mathrm{SU}} = \mathbf{X}^{-1} \mathbf{z}_{\mathrm{SU}} = \mathbf{j}_{\mathrm{SU}} + \mathbf{X}^{-1} \mathbf{n}_{s}.
\end{equation}

The estimation error is defined by the term $\mathbf{X}^{-1} \mathbf{n}_{s}$. Consequently, the sensing MSE is derived as:
\begin{equation}\label{SU sensing error}
    \epsilon_{\mathrm{SU}} = \mathbb{E}\left[ \left| \mathbf{X}^{-1} \mathbf{n}_{s} \right|^{2} \right] \geq \mathrm{Tr}\left(\sigma_s^2 \left( \mathbf{Q} \mathbf{R}_{x} \mathbf{Q}^{\mathrm{H}} \right)^{-1} \right),
\end{equation}
where $\mathbf{Q}$ is the diagonal weight matrix previously defined, and the inequality comes from Jensen's inequality.

\begin{remark}\label{remark 1}
    Comparing \eqref{cu Sensing error} and \eqref{SU sensing error}, it can be observed that the proposed design creates a sensing performance gap in favor of the SU, such that the sensing error at the CU is larger than that at the SU. In the absence of AN, i.e., when $\boldsymbol{\Sigma} =\mathbf{0}$, \eqref{cu Sensing error} reduces to \eqref{SU sensing error} since $\mathbf{R}_{x}=\mathbf{R}_{\tilde{w}}$. Therefore, imposing $\boldsymbol{\Sigma} \succeq \boldsymbol{\eta}$ for some threshold matrix $\boldsymbol{\eta} \succ \mathbf{0}$, serves as a sufficient condition to preserve the sensing advantage of the SU over the CU.
\end{remark}

\subsubsection{Unauthorized Service}

When acting as a communication Eve, the SU performs CP removal and an $N_c$-point FFT. Since the AN precoder $\mathbf{U}$ is constructed based on the null space of $\mathbf{H}_{\mathrm{CU}}$, the non-orthogonality between $\mathbf{U}$ and the SU channel $\mathbf{H}_{\mathrm{SU}}$ ensures that AN is injected into the SU received signal. The resulting decoding SINR at the SU is:
\begin{equation}\label{SU SINR}
    \gamma_{\mathrm{SU, decode}} = \frac{\mathrm{Tr}\left( \tilde{\mathbf{H}}_{\mathrm{SU}}\mathbf{W} \tilde{\mathbf{H}}_{\mathrm{SU}}^{\mathrm{H}} \right)}{\mathrm{Tr}\left( \mathbf{F} \mathbf{R}_{\mathrm{CP}} \mathbf{H}_{\mathrm{SU}} \mathbf{U} \boldsymbol{\Sigma} \mathbf{U}^{\mathrm{H}} \mathbf{H}_{\mathrm{SU}}^{\mathrm{H}} \mathbf{R}_{\mathrm{CP}}^{\mathrm{H}} \mathbf{F}^{\mathrm{H}}  \right) + \sigma^2}.
\end{equation}

Other data security metrics, such as the secrecy rate, may also be adopted. However, since the secrecy rate is fundamentally determined by the underlying SNR/SINR, we use \eqref{cu snr} and \eqref{SU SINR} as design metrics to guarantee data security while reducing the design complexity.

\section{Dual-Secured ISAC Design}

This section formulates the optimization problem of the proposed dual-secure design and presents the solution.

\subsection{Problem Formulation}
In this work, the objective is to maximize the performance of the legitimate services for both the CU and SU. Specifically, this corresponds to maximizing the communication SNR at the CU while minimizing the sensing error at the SU. Security constraints are imposed subject to the power budget. Accordingly, the optimization problem is formulated as:
\begin{subequations}\label{opt 1}
\begin{align}
    \max_{\mathbf{W} \succeq 0, \boldsymbol{\Sigma} \succeq 0} & \quad \kappa_1 \gamma_{\mathrm{CU}} - \kappa_2 \epsilon_{\mathrm{SU}} \label{opt1 obj} \\
    \text{s.t.} \quad & \quad \gamma_{\mathrm{SU, decode}} \leq \eta_{b}, \label{opt1 b} \\
    & \quad \epsilon_{\mathrm{CU}} \geq \eta_{c}, \label{opt1 c} \\
    & \quad \mathrm{Tr}\left( \mathbf{W} \right) + \mathrm{Tr}\left( \boldsymbol{\Sigma} \right) \leq P_{t}, \label{opt1 d} 
\end{align}
\end{subequations}
where $\eta_{b,c}$ are threshold values, $\kappa_{1,2}$ are trade-off terms including scaling factors, and $P_t$ is the power budget.

\subsection{Problem Transformation}

In \eqref{opt 1}, the term $\epsilon_{\mathrm{SU}}$ in the objective function and constraint \eqref{opt1 c} are non-convex. The remaining objective function and constraints are convex.

To first handle the minimization of $\epsilon_{\mathrm{SU}}$, we employ the Schur complement \cite{naghibi2011mimo}. By introducing an auxiliary matrix $\mathbf{M} \succeq \mathbf{0}$, the term $-\kappa_2 \epsilon_{\mathrm{SU}}$ in the objective function can be equivalently reformulated as $-\kappa_2 \mathrm{Tr}(\mathbf{M})$, together with the constraint
\begin{equation}\label{schur}
    \begin{bmatrix}
        \mathbf{M} & \mathbf{I}_{N_c + N_{cp}} \\
        \mathbf{I}_{N_c + N_{cp}} & \frac{1}{\sigma_s^2}\mathbf{Q}^{\mathrm{H}}\mathbf{R}_{x}\mathbf{Q}
    \end{bmatrix} \succeq \mathbf{0}.
\end{equation}

Constraint \eqref{opt1 c} may be handled using the techniques in \cite{sun2016majorization}, such as cyclic minimization or Taylor expansion with alternating optimization. However, these approaches are inherently iterative and often sensitive to the initialization. To improve computational tractability, we instead adopt a trace-based relaxation. Specifically, for $\mathbf{A} \succeq \mathbf{0}$ and $\mathbf{B} \succ \mathbf{0}$ with $\mathbf{A}, \mathbf{B} \in \mathbb{C}^{N \times N}$, the condition
\begin{equation}\label{trace_sufficient}
    \mathbf{B}^{-1/2}\mathbf{A}\mathbf{B}^{-1/2} \succeq \frac{\eta}{N}\mathbf{I},
\end{equation}
is sufficient for
\begin{equation}
    \mathrm{Tr}\!\left(\mathbf{A}\mathbf{B}^{-1}\right) \ge \eta.
\end{equation}

Accordingly, \eqref{opt1 d} is conservatively replaced by the following convex matrix inequality:
\begin{equation}\label{trace inequality}
    \left( \sigma_j^2 \mathbf{Q} \mathbf{U} \boldsymbol{\Sigma} \mathbf{U}^{\mathrm{H}} \mathbf{Q}^{\mathrm{H}} + \sigma_s^2 \mathbf{I} \right)
    - \frac{\eta_c}{N_c + N_{cp}}\left( \mathbf{Q}^{\mathrm{H}}\mathbf{R}_{\tilde{w}}\mathbf{Q} \right)
    \succeq \mathbf{0}.
\end{equation}

As a result, \eqref{opt 1} is reformulated as
\begin{subequations}\label{opt 2}
\begin{align}
    \max_{\mathbf{W} \succeq 0, \boldsymbol{\Sigma}\succeq 0, \mathbf{M} \succeq 0} \quad
    & \kappa_1 \gamma_{\mathrm{CU}} - \kappa_2 \mathrm{Tr}\left(\mathbf{M}\right) \label{opt2 obj} \\
    \text{s.t.} \hspace{1.2cm} & \eqref{opt1 b},~\eqref{opt1 d},~\eqref{schur},~\eqref{trace inequality}.
\end{align}
\end{subequations}

Problem \eqref{opt 2} is convex and can therefore be solved to global optimality using standard convex optimization solvers.

\section{Simulation Results}

In this section, we evaluate the proposed dual-secure AN design by analyzing the trade-off between performance metrics. We consider $N_{c} = 64$, $N_{cp} = 16$, and all channels have maximum of $8$ taps. Unless specified, the noise power is $-70$ dBm, and the power budget $P_{t}=15$ dBm. The data symbols are drawn randomly from the 4-quadrant amplitude-modulated (4-QAM) constellation. Finally, in \eqref{opt 2}, the trade-off coefficients are set to half to balance the services provided to legitimate users. All results are averaged over 300 Monte Carlo runs.

\subsection{Communication Performance}

\begin{figure}[t]
    \centering
    \includegraphics[width=0.75\columnwidth]{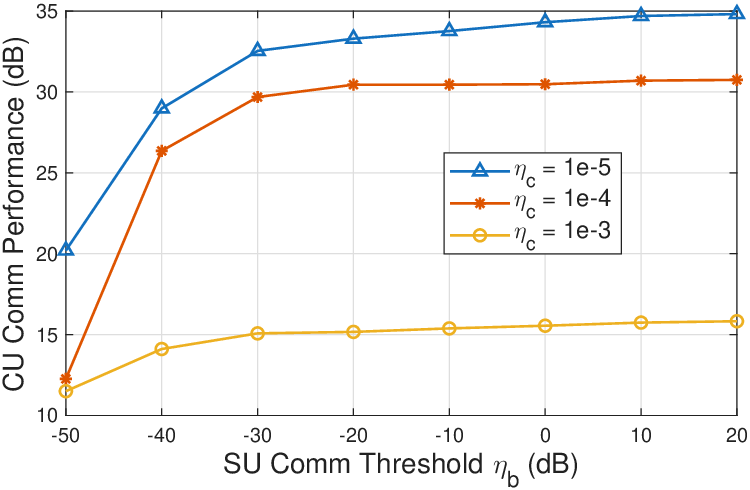}
    \caption{The CU communication performance, versus SU decoding SINR threshold $\eta_b$, under different CU sensing MSE thresholds $\eta_c$.}
    \label{fig comm secure}
\end{figure}

\begin{figure}[t]
    \centering
    \includegraphics[width=0.75\columnwidth]{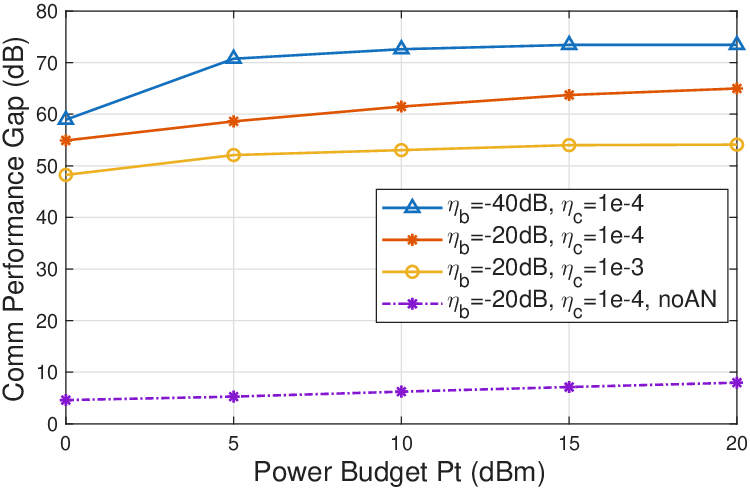}
    \caption{The communication performance gap versus transmit power budget $P_t$ under different pairs of SU decoding SINR threshold $\eta_b$ and CU sensing MSE thresholds $\eta_c$.}
    \label{fig comm gap}
\end{figure}

Fig.~\ref{fig comm secure} illustrates the CU communication performance, measured by the per-subcarrier SNR, versus the SU decoding SINR upper bound $\eta_b$ under different CU sensing MSE lower bounds $\eta_c$. As $\eta_b$ decreases, stricter communication security is imposed, requiring more power to be allocated to AN to suppress the SU decoding capability. Consequently, less power is available for the information-bearing signal, leading to a reduction in the QoS of CU. Moreover, a larger $\eta_c$ leads to a higher sensing error at the CU, further reducing its QoS, as additional AN power allocation is required to impair the CU's sensing capability.

Fig.~\ref{fig comm gap} shows the performance gap versus the transmit power budget \(P_t\) under different pairs of \((\eta_b,\eta_c)\). The performance gap is defined as the difference between the per-subcarrier SNR of the CU and the per-subcarrier SINR of the SU, and thus reflects the achieved level of communication security. As the transmit power budget increases, the performance gap increases and gradually saturates, indicating that additional transmit power enhances communication security but diminishes returns. Moreover, reducing \(\eta_b\) enlarges the performance gap, since a more stringent communication security requirement requires stronger AN, which suppresses the SU decoding SINR more significantly while preserving the CU SNR. In contrast, increasing \(\eta_c\) reduces the performance gap. This is because a more stringent sensing security requirement causes a larger portion of the available power to be allocated to AN rather than to improving the CU communication performance. Compared with the non-secure design in \cite{marzban2018securing}, denoted by ``noAN'', the proposed design achieves a performance gap that is at least five times larger. This improvement stems from the fact that the non-secure design focuses solely on enhancing the CU communication performance without explicitly limiting the decoding capability of the SU. Consequently, the SU can also benefit from the increased transmit power, resulting in only a limited performance gap.

\subsection{Sensing Performance}

\begin{figure}[t]
    \centering
    \includegraphics[width=0.75\columnwidth]{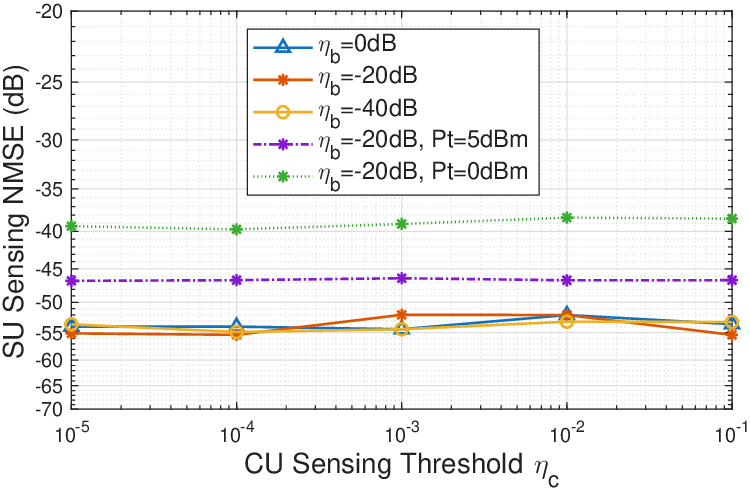}
    \caption{The SU channel estimation NMSE, versus CU sensing MSE threshold $\eta_c$, under different SU decoding SINR threshold $\eta_b$ and power budget $P_t$.}
    \label{fig sensing security}
\end{figure}

\begin{figure}[t]
    \centering
    \includegraphics[width=0.75\columnwidth]{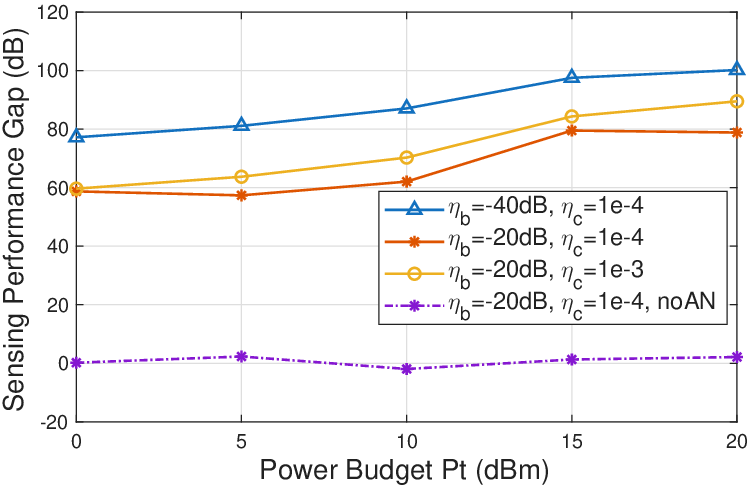}
    \caption{The sensing performance gap versus transmit power budget $P_t$ under different pairs of SU decoding SINR threshold $\eta_b$ and CU sensing MSE threshold $\eta_c$.}
    \label{fig sensing gap}
\end{figure}

From \eqref{SU sensing error}, the SU sensing performance depends only on the total transmit power and is independent of the power allocation between the AN and the data-bearing signal. This is confirmed in Fig.~\ref{fig sensing security}, where the SU sensing normalized mean squared error (NMSE) remains nearly unchanged as $\eta_c$ varies, while increasing as the total transmit power decreases. In addition, increasing $\eta_b$ raises the CU sensing MSE by allocating more power to AN to increase CU sensing mismatch, which does not affect the SU sensing performance.

Fig.~\ref{fig sensing gap} illustrates the sensing performance gap, defined as the NMSE of CU minus the NMSE of SU, which characterizes sensing security by quantifying the sensing error difference. As the transmit power budget increases, the sensing performance gap increases only slightly. This is because the CU sensing NMSE remains substantially higher than that of the SU over the considered power range, and hence further increases in transmit power provide only a limited gain. Reducing \(\eta_c\) decreases the sensing performance gap, since a less stringent sensing security requirement requires less AN to degrade the CU sensing performance. Furthermore, reducing \(\eta_b\) improves communication security and simultaneously improves sensing security, as both metrics are fundamentally tied to the improved allocation of AN power. Compared with the non-secure benchmark design, the sensing performance gap is nearly zero, indicating that the CU and SU experience comparable sensing errors, which enables unauthorized acquisition of sensing information.

\section{Conclusion and Future Work}
This paper investigates data eavesdropping and unauthorized sensing in an indoor OFDM-ISAC system, where the sensing task is target impulse response estimation. To address this dual-security problem, a temporal AN scheme is proposed by embedding AN into the CP of the OFDM waveform. The AN can be completely removed at the CU using standard OFDM receiver processing, thereby preserving its communication performance, while remaining non-removable at the SU, thus preventing data eavesdropping. Meanwhile, the embedded AN is also present in the target echo received by both users. Since only the SU receives the AN, this signal alignment enables the SU to accurately estimate the target response. By contrast, because the CU does not receive the AN, a mismatch arises during target impulse response estimation, leading to severe performance degradation and thereby ensuring sensing security. Simulation results verify the effectiveness of the proposed design.

Future work may consider more practical scenarios involving imperfect CSI, indoor channel modeling in rich-scattering environments, imperfect recovery of the reference signals, and stronger adversaries capable of estimating not only the channel but also the artificial noise.

\bibliographystyle{IEEEtran}
\bibliography{ref}

\end{document}